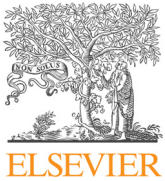
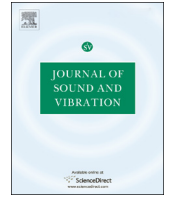
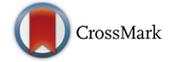

# Structures vibration control via Tuned Mass Dampers using a co-evolution Coral Reefs Optimization algorithm

S. Salcedo-Sanz [a],[*], C. Camacho-Gómez [a], A. Magdaleno [b], E. Pereira [a], A. Lorenzana [b]

[a] Department of Signal Processing and Communications, Universidad de Alcalá, Madrid, Spain
[b] ITAP, EII, Universidad de Valladolid, Valladolid, Spain

ABSTRACT

In this paper we tackle a problem of optimal design and location of Tuned Mass Dampers (TMDs) for structures subjected to earthquake ground motions, using a novel meta-heuristic algorithm. Specifically, the Coral Reefs Optimization (CRO) with Substrate Layer (CRO-SL) is proposed as a competitive co-evolution algorithm with different exploration procedures within a single population of solutions. The proposed approach is able to solve the TMD design and location problem, by exploiting the combination of different types of searching mechanisms. This promotes a powerful evolutionary-like algorithm for optimization problems, which is shown to be very effective in this particular problem of TMDs tuning. The proposed algorithm's performance has been evaluated and compared with several reference algorithms in two building models with two and four floors, respectively.

© 2017 Elsevier Ltd. All rights reserved.

## 1. Introduction

Problems in structural optimization are often characterized by search spaces of extremely high dimensionality and nonlinear objective functions [1]. In these optimization problems, classical approaches do not lead, in general, to good solutions, or in many occasions they are just not applicable, due to the unmanageable search space structure or its huge size, which implies an extremely high computation cost. In this context, modern optimization meta-heuristics have been successfully applied to an important number of structural optimization problems [2]. Meta-heuristics algorithms have been shown as a possibility to obtain a *good enough* solution to a given problem which cannot be tackled with exact algorithms.

There are different meta-heuristics that have been applied to structural engineering problems. Genetic and evolutionary algorithms [3] have been applied to the optimization of discrete structures in [4]. There have been other works that applied genetic algorithms in structural optimization problems such as shape optimization [5], optimization of 3D trusses [6], impact load characterization of concrete structure [7], the plane stress problem [8] or welded beam optimization problems [9]. The particle swarm optimization algorithm [10] is another important meta-heuristic which has been successfully applied to structural optimization problems, such as truss layout [11] or truss structures optimization [12]. The Harmony Search approach [13,1] and the teaching-based learning algorithm [14–16] have also been used to solve mechanical design optimization problems. In the last few years, alternative modern meta-heuristics based on physics process have been applied to structural optimization problems, such as the Big-Bang Big-Crunch algorithm [17], the colliding bodies optimization algorithm [18], the Ray optimization [19] or the charged system search algorithm [20].

In this paper, a novel co-evolution meta-heuristic, the Coral Reefs Optimization algorithm with Substrate Layer (CRO-SL) [21], is applied to the design and location of Tuned Mass Dampers (TMDs) for structures subjected to earthquake ground motions. A TMD, which can be used for passive and semi-active control strategies, improves the vibration response of a structure by increasing its damping (i.e. energy dissipation) and/or stiffness (i.e. energy storage) through the application of forces generated in response to the movement of the structure [22]. In the case of structures with spatially distributed and closely spaced natural frequencies, the TMD design may not be obvious, because Den Hartog's theory [23] cannot be applied due to the existence of a coupling between the motions of the vibration modes of the structures and the used TMDs [24]. Multi-storey buildings are good examples of structures with spatially distributed and closely spaced natural frequencies. For example, Greco et al. [25] proposes a robust optimum design of tuned mass dampers installed on multi-degree-of-freedom systems subjected to stochastic seismic actions. Other similar examples can be found in [26] and [27]. In this work, the generalized framework presented in [28] is used to formulate a $N$ floor building where $M$ TMDs must be installed. Unlike [28], where the position of each TMD is fixed (p.e., $M$ TMDs in one floor or one TMD for each floor), this work proposes a modification that allows the optimization algorithm deciding the position of each TMD (i.e., a TMD can be placed at any floor to damp any vibration mode). In addition, an interval for the mass, damping and stiffness is defined for each TMD. Thus, the optimization algorithm will try to find the best solution by obtaining the $4 \times M$ parameters (3 physical parameters and the TMD location). As previously mentioned, in this paper the optimization algorithm proposed is a co-evolution approach, the CRO-SL algorithm, which is able to combine several types of searching mechanisms into just one population structure, obtaining a powerful evolutionary-like algorithm for optimization problems.

The structure of the remainder of the paper is as follows: next section describes the generalized framework used to obtain the optimal design and location for TMDs installed on a $N$ storey building. Section 3 presents the main characteristics of the original CRO, including the different operators and the algorithm's dynamics. Section 3 describes the proposed CRO-SL version, including the definition of *substrate layer*, and, in this case, how it represents the co-evolution of different searching mechanism with the rules of the CRO. Section 4 presents the computational evaluation section, where the proposed algorithm's performance is evaluated and compared with a reference algorithm. The CRO-SL application is validated in Section 5, where a experimental set-up is used to test the optimum TMD design (location and parameters) obtaining by CRO-SL. Section 6 closes the paper by giving some final conclusions and remarks on this research.

## 2. Problem definition

The $N$ storey building can be modelled as a $N$ degree of freedom system (see Fig. 1 (a)), where the mass is concentrated at each floor ($m_1, m_2, ..., m_N$), $k_i$ and $c_i$ are, respectively, the $i^{th}$ floor stiffness and damping coefficient (relative to $(i-1)^{th}$ floor or to the ground if $i=1$).

If the applied forces in each floor $\left(\mathbf{f} = [f_1, f_2, ..., f_N]^T\right)$ and the acceleration of the ground ($a_g$) are considered as inputs, the differential equation of the building can be represented as follows:

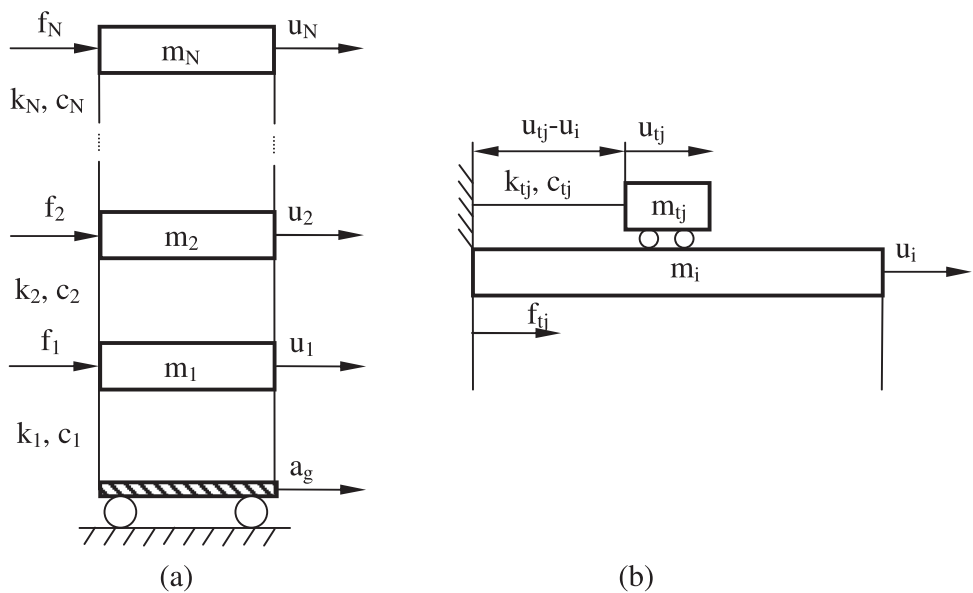

**Fig. 1.** (a) N storey building and (b) TMD models.

$$\mathbf{M}\ddot{\mathbf{u}} + \mathbf{C}\dot{\mathbf{u}} + \mathbf{K}\mathbf{u} = -\mathbf{M}\mathbf{r}_g a_g + \mathbf{f}, \tag{1}$$

where $\mathbf{M} = diag(m_1, m_2, \ldots, m_N)$ is the structural mass matrix,

$$\mathbf{K} = \begin{bmatrix} k_1 + k_2 & -k_2 & 0 & \cdots & 0 \\ -k_2 & k_2 + k_3 & -k_3 & \cdots & 0 \\ \vdots & \vdots & \ddots & \cdots & \vdots \\ 0 & 0 & -k_{N-1} & k_{N-1} + k_N & -k_N \\ 0 & 0 & 0 & -k_N & k_N \end{bmatrix} \tag{2}$$

is the structural stiffness matrix, $\mathbf{C}$ is the proportional damping matrix, $\mathbf{u} = [u_1, u_2, \ldots, u_N]^T$ are the floor displacements relative to the base and $\mathbf{r}_g = [1\cdots 1]^T$ is the influence vector of the ground acceleration. In order to simplify the model, the damping matrix is assumed as $\mathbf{C} = \frac{2\xi_s \omega_1 \omega_2}{\omega_1 + \omega_2}\mathbf{M} + \frac{2\xi_s}{\omega_1 + \omega_2}\mathbf{K}$. Note that (like at reference [28]), the dissipation matrix assuming that damping forces depend only on generalized velocities is not the only linear model of vibration damping (see reference [29] for a detailed discussion). However, for simplification of the illustration, a dissipation matrix damping is assumed for the shear building model (this does not affect the generality of the framework, because the performance measures are formulated using the plant without any assumptions rather than linearity).

The state space state model can be deduced from Eq. (1)

$$\begin{bmatrix} \dot{\mathbf{u}} \\ \ddot{\mathbf{u}} \end{bmatrix} = \begin{bmatrix} \mathbf{0}_{N \times N} & \mathbf{I}_{N \times N} \\ -\mathbf{M}^{-1}\mathbf{K} & -\mathbf{M}^{-1}\mathbf{C} \end{bmatrix} \begin{bmatrix} \mathbf{u} \\ \dot{\mathbf{u}} \end{bmatrix} + \begin{bmatrix} \mathbf{0}_{N \times 1} & \mathbf{0}_{N \times N} \\ -\mathbf{r}_g & \mathbf{M}^{-1} \end{bmatrix} \begin{bmatrix} a_g \\ \mathbf{f} \end{bmatrix}, \tag{3}$$

$$\mathbf{y} = \ddot{\mathbf{u}} + \mathbf{r}_g a_g = \begin{bmatrix} -\mathbf{M}^{-1}\mathbf{K} & -\mathbf{M}^{-1}\mathbf{C} \end{bmatrix} \begin{bmatrix} \mathbf{u} \\ \dot{\mathbf{u}} \end{bmatrix} + \mathbf{M}^{-1}\mathbf{f},$$

where $\mathbf{y}$ is the vector formed by the absolute accelerations (i.e., the accelerations measured with the accelerometers installed at each floor).

The TMD can be modelled as a one degree of freedom system (see Fig. 1 (b)), where $m_{tj}$ is the mass, $k_{tj}$ and $c_{tj}$ are the TMD linear stiffness and damping coefficient of the $j^{th}$ TMD relative to the $i^{th}$ floor. The differential equation of the TMD relates the accelerations of the ground floor ($a_g$) the mass of the $i^{th}$ floor ($u_i$) and the mass of the $j^{th}$ TMD as follows:

$$m_{t,j}\ddot{u}_{tj} + c_{t,j}\dot{u}_{t,j} + k_{t,j}u_{t,j} - c_{t,j}\dot{u}_i - k_{t,j}u_i = -m_{t,j}a_g, \tag{4}$$

where the force exerted by the $j^{th}$ TMD on the $i^{th}$ floor is:

$$f_{t,j} = k_{t,j}(u_{t,j} - u_i) + c_{t,j}(\dot{u}_{t,j} - \dot{u}_i) = k_{t,j}u_{r,ij} + c_{t,j}\dot{u}_{r,ij}, \tag{5}$$

and the relative displacement between $j^{th}$ TMD and $i^{th}$ floor is defined as $u_{r,ij} = u_{t,j} - u_i$. If the variable $u_{r,ij}$ is considered, the Eq. (4) can be arranged as the following:

$$m_{t,j}\ddot{u}_{r,ij} + c_{t,j}\dot{u}_{r,ij} + k_{t,j}u_{r,ij} = -m_{t,j}(a_g + \ddot{u}_i) = -m_{t,j}y_i. \tag{6}$$

Note that Eqs. (3), (5) and (6) define the system formed by the $N$ floor building and the $M$ TMDs, which can be re-presented as in Fig. 2. Note also that the values of the applied forces in each floor ($f_i$) are equal to the sum of the forces of all TMD located in this floor according to Eq. (5). The optimization problem proposed in this work consists of minimizing the maximum of the Frequency Response Functions (FRFs) defined between each output ($y_i$) and the ground acceleration ($a_g$). Thus, the Eqs. (3), (5) and (6) must be defined at frequency domain. First of all, the state space model of Eq. (3) can be defined as the following $N \times (N + 1)$ matrix of transfer functions [30]:

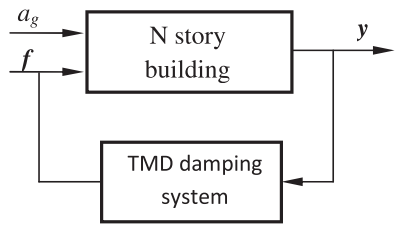

Fig. 2. General framework.

$$\begin{bmatrix} Y_1(s) \\ Y_2(s) \\ \vdots \\ Y_N(s) \end{bmatrix} = \begin{bmatrix} G_{11}(s) & G_{12}(s) & \cdots & G_{1N}(s) & G_{1g}(s) \\ G_{21}(s) & G_{22}(s) & \cdots & G_{2N}(s) & G_{2g}(s) \\ \vdots & \vdots & \ddots & \vdots & \vdots \\ G_{N1}(s) & G_{N2}(s) & \cdots & G_{NN}(s) & G_{Ng}(s) \end{bmatrix} \begin{bmatrix} F_1(s) \\ F_2(s) \\ \vdots \\ F_N(s) \\ A_g(s), \end{bmatrix} \quad (7)$$

where $s$ is the (complex) frequency variable and the capital letters $Y_i(s)$, $F_i(s)$ and $A_g(s)$ denote the Laplace transform of $y_i$, $f_i$ and $a_g$, respectively. Finally, $G_{i_y i_f}(s)$ is the transfer function between the acceleration measured at $i_y^{th}$ floor and the force applied to the $i_f^{th}$ floor and $G_{i_y g}(s)$ is the transfer function between the acceleration measured at $i_y^{th}$ and the ground acceleration ($a_g$).

From Eqs. (5) and (6) the following transfer function ($H_j(s)$) for the TMD system can be deduced:

$$\frac{F_{t,j}(s)}{Y_i(s)} = -\frac{m_{t,j}(c_{t,j}s + k_{t,j})}{m_{t,j}s^2 + c_{t,j}s + k_{t,j}} = -m_{t,j}\frac{2\xi_{t,j}\omega_{t,j}s + \omega_{t,j}^2}{s^2 + 2\xi_{t,j}\omega_{t,j}s + \omega_{t,j}^2} = H_j(s), \quad (8)$$

where $\omega_{t,j} = \sqrt{(k_{t,j}/m_{t,j})}$ and $\xi_{t,j} = c_{t,j}/2\sqrt{k_{t,j}m_{t,j}}$ are the natural frequency and damping ratio of the $j^{th}$ TMD as an isolated system, respectively, and $F_{t,j}(s)$ and $Y_i(s)$ are the Laplace transform of $f_{t,j}$ and $y_i$.

The Eqs. (7) and (8) can be connected as in the general framework of Fig. 2 with the following equation:

$$F_i(s) = \sum_{j=1}^M K_{j,i} H_j(s) Y_i(s), \quad (9)$$

where $K_{ji} = 1$ if the TMD $j$ is placed on $i^{th}$ floor. Once the general framework is defined in the Laplace domain, the optimization problem consist of minimizing the following functional:

$$g(\mathbf{x}) = \max\left\{ \left\|\frac{Y_1(j\omega)}{A_g(j\omega)}\right\|_\infty, \left\|\frac{Y_2(j\omega)}{A_g(j\omega)}\right\|_\infty, \ldots, \left\|\frac{Y_N(j\omega)}{A_g(j\omega)}\right\|_\infty \right\}, \quad (10)$$

by finding the optimal parameters of $\mathbf{x} = [\mathbf{\Omega}_t, \mathbf{\Xi}_t, \mathbf{M}_t, \mathbf{FB}]$, where $\|\|_\infty$ is the infinity norm, $\mathbf{\Omega}_t = [\omega_{t,1}, \omega_{t,2}, \ldots, \omega_{t,M}]$, $\mathbf{\Xi}_t = [\xi_{t,1}, \xi_{t,2}, \ldots, \xi_{t,M}]$, $\mathbf{M}_t = [m_{t,1}, m_{t,2}, \ldots, m_{t,M}]$ and $\mathbf{FB} = [fb_1, fb_2, \ldots, fb_M]$. Note that the parameter $fb_j = i$ if the TMD $j$ is placed on the $i^{th}$ floor (i.e., $K_{ji} = 1$). This problem can be formulated as follows:

$$\min_{\mathbf{x}}(g(\mathbf{x})). \quad (11)$$

## 3. The Coral Reefs Optimization algorithm with substrate layer

This section presents the CRO-SL proposed in this paper for tackling a problem of design and location of TMDs. First, we present the basic CRO algorithm, which will be modified with a *substrate layer* in order to obtain a competitive co-evolution algorithm with different exploration procedures.

### 3.1. Basic CRO

The CRO is an evolutionary computation meta-heuristic for optimization, recently proposed in [31], which is based on simulating the corals' reproduction and coral reefs' formation processes. It has been successfully applied to a number of different applications and optimization problems [32–37]. Basically, the CRO is based on the artificial modeling of a coral reef $\mathcal{R}$, consisting of a $n \times m$ grid. We assume that each square $(i,j)$ of $\mathcal{R}$ is able to allocate a coral $\tau_{ij}$ (candidate solution to the problem, called as $\mathbf{x}$ in the problem's statement above). The CRO algorithm is first initialized at random by assigning some squares in $\mathcal{R}$ to be occupied by corals (i.e. solutions to the problem) and some other squares in the grid to be empty, i.e. holes in the reef where new corals can freely settle and grow in the future. The rate between free/occupied squares in $\mathcal{R}$ at the beginning of the algorithm is denoted as $\rho \in \mathbb{R}(0, 1)$ and referred to as initial occupation factor. Each coral is labeled with an associated *health* function $f(\tau_{ij}): \mathcal{A} \to \mathbb{R}$ that corresponds to the problem's objective function. The CRO is based on the fact that the reef will evolve and develop as long as healthier or stronger corals (which represent better solutions to the problem at hand) survive, while less healthy corals perish.

After the reef initialization described above, the phase of reef formation is artificially simulated. This phase consists of $\alpha$ iterations: at each of such iterations the corals' reproduction in the reef is emulated by applying different operators and processes as described in Algorithm 1: a modeling of corals' sexual reproduction (broadcast spawning and brooding). After the reproduction stage, the set of formed larvae (namely, newly produced solutions to the problem) attempts to find a place on the reef to develop and further reproduce. This deployment may occur in a free space inside the reef (hole), or in an occupied location, by fighting against the coral currently settled in that place. If larvae are not successful in locating a place to settle after a number of attempts, they are considered as preyed by animals in the reef. The coral builds a new reef layer in every iteration.

**Algorithm 1.** Pseudo-code for the CRO algorithm.

```
Require: Valid values for the parameters controlling the CRO algorithm
Ensure: A single feasible individual with optimal value of its fitness
1:    Initialize the algorithm
2:    for each iteration of the simulation do
3:        Update values of influential variables: predation probability, etc.
4:        Sexual reproduction processes (broadcast spawning and brooding)
5:        Settlement of new corals
6:        Predation process
7:        Evaluate the new population in the coral reef
8:    end for
9:    Return the best individual (final solution) from the reef
```

We detail here the specific definition of the different operators that form the classical CRO algorithm:

1. **Sexual reproduction**: The CRO model implements two different kinds of sexual reproduction: external and internal.
   (a) **External sexual reproduction** or *broadcast spawning*: the corals eject their gametes to the water, from which male-female couples meet and combine together to produce a new larva by sexual crossover. In Nature, some species are able to combine their gametes to generate mixed polyps even though they are different from each other. In the CRO algorithm, external sexual reproduction is applied to a usually high fraction $F_b$ of the corals. The couple selection can be done uniformly at random or by resorting to any fitness proportionate selection approach (e.g. roulette wheel). In the original version of the CRO, standard crossover (one point or two-points) are applied in the broadcast spawning process.
   (b) **Internal sexual reproduction** or *brooding*: CRO applies this method to a fraction $(1 - F_b)$ of the corals in the reef. The brooding process consists of the formation of a coral larva by means of a random mutation of the brooding-re-productive coral (self-fertilization considering hermaphrodite corals). The produced larvae is then released out to the water in a similar fashion than that of the larvae generated through broadcast spawning.
2. **Larvae settlement**: once all larvae are formed at iteration $k$ through reproduction, they try to settle down and grow in the reef. Each larva will randomly attempt at setting in a square $(i,j)$ of the reef. If the location is empty (free space in the reef), the coral grows therein no matter the value of its health function. By contrast, if another coral is already occupying the square at hand, the new larva will set only if its health function is better than the fitness of the existing coral. We define a number of attempts $\mathcal{N}_{att}$ for a larva to set in the reef: after $\mathcal{N}_{att}$ unsuccessful tries, it will not survive to following iteration.
3. **Depredation**: corals may die during the reef formation phase of the reef. At the end of each iteration, a small number of corals can be preyed, thus liberating space in the reef for the next iteration. The depredation operator is applied under a very small probability $P_d$, and exclusively to a fraction $F_d$ of the worse health corals.

### 3.2. CRO-SL

The original CRO algorithm is based on the main processes of coral reproduction and reef formation that occur in nature. However, there are many more interactions in real reef ecosystem that can be also modelled and incorporated to the CRO approach to improve it. For example, different studies have shown that successful recruitment in coral reefs (i.e., successful settlement and subsequent survival of larvae) depends on the type of substrate on which they fall after the reproduction process [38]. This specific characteristic of coral reefs was first included in the CRO in [39], in order to solve different instances of the Model Type Selection Problem for energy applications. The CRO with substrates is a general approach: it can be defined as an algorithm for competitive co-evolution, where each substrate layer represents different processes (different models, operators, parameters, constraints, repairing functions, etc.). This idea of CRO with substrate layers, was extended as a fully competitive co-evolution search mechanism in [21], where each substrate layer represents a different exploration mechanism. In [40] the interested reader can find more details on alternative co-evolution versions of the CRO algorithm. In this section we describe the main ideas of the CRO-SL as co-evolution search algorithm.

The inclusion of substrate layers in the CRO can be done, in a general way, in a straightforward manner: we redefine the artificial reef considered in the CRO in such a way that each cell of the square grid $\mathcal{R}$ representing the reef is now defined by 3 indexes $(i, j, t)$, where $i$ and $j$ stand for the cell location in the grid, and index $t \in T$ defines the substrate layer, by indicating which structure (model, operator, parameter, etc.) is associated with the cell $(i,j)$. Each coral in the reef is then processed in a different way depending on the specific substrate layer in which it falls after the reproduction process. Note that this modification of the basic algorithm does not imply any change in the corals' encoding. When the CRO-SL is focused on improving the searching capabilities of the classical CRO approach, each substrate layer is defined as a different implementation of an exploration procedure. Thus, each coral will be processed in a different way in the reproduction step of the algorithm. Fig. 3 shows an example of the CRO-SL, with five different substrate layers. Each one is assigned to a different exploration process, Harmony Search based, Differential Evolution, Gaussian Mutation, M-Points Crossover or 2-Points Crossover. Of course this is only an example and any other distribution of search procedures can be defined in the algorithm. In the specific CRO-SL tested in this paper, each substrate layer only affects to the calculation of the larvae coming from the broadcast spawning process, whereas we have considered the same brooding procedure for all the corals in the reef.

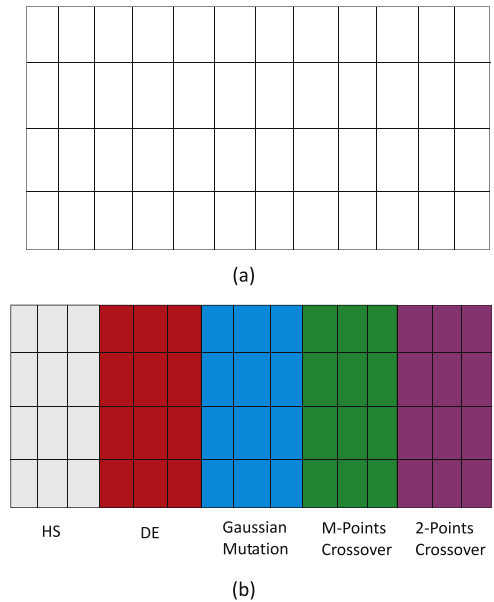

**Fig. 3.** Example of CRO-SL and comparison with the original reef in the CRO; (a) Reef considered in the original CRO; (b) Reef in the CRO-SL, where five substrate layers associated with the broadcast spawning process have been considered (Harmony Search (HS), Differential Evolution (DE), Gaussian Mutation, M-Points Crossover and 2-Points Crossover).

There are some important remarks that can be done regarding the CRO-SL approach. First, note that the original CRO is a meta-heuristic based on exploitation of solutions, and leaves the specific exploration open (in the same manner as, for example, Simulated Annealing [41]). This way, the CRO-SL can be seen as a generalization of the original CRO, that does not modify the dynamics of the algorithm (so it can be still outlined following Algorithm 1). The only difference is the specific implementation of the broadcast spawning procedure, which now depends on the specific substrate to which the coral is associated. Second, as has been previously mentioned, the CRO-SL can be seen as a competitive co-evolution procedure. The CRO-SL is a general procedure to co-evolve different models, operators, parameter values, etc., with the only requisite that there is only one health function defined in the algorithm. In this sense, note that the CRO-SL makes a competitive co-evolution of different searching models or patterns within one population of solutions.

## 4. Computational evaluation and comparisons

The examples carried out to evaluate the proposed CRO-SL in this context consist of designing and locating $M$ TMDs on a $N$ floor building. The CRO-SL parameters used in the experiments are shown in Table 1.

Five different substrates layers have been used in the experiments carried out:

1. HS: Mutation from the Harmony Search algorithm.
2. DE: Mutation from Differential Evolution algorithm (with $F=0.6$).
3. 2Px: Classical 2-points crossover.
4. GM: Gaussian Mutation, with a $\delta$ value linearly decreasing during the run, from $0.2 \cdot (A - B)$ to $0.02 \cdot (A - B)$, where $[B, A]$ is the domain search.
5. MPx: Multi-points crossover.

**Table 1**
Parameters values used in the hybridization of the CRO-SL.

| Parameter | Description | Value |
| --- | --- | --- |
| Reef | Reef size | 120 |
| $F_b$ | Frequency of broadcast spawning | 97% |
| $N_{att}$ | Number of tries for larvae settlement | 3 |
| $P_d$ | Probability of depredation | 5% |
| $\alpha$ | Maximum number of iterations | 1000 |

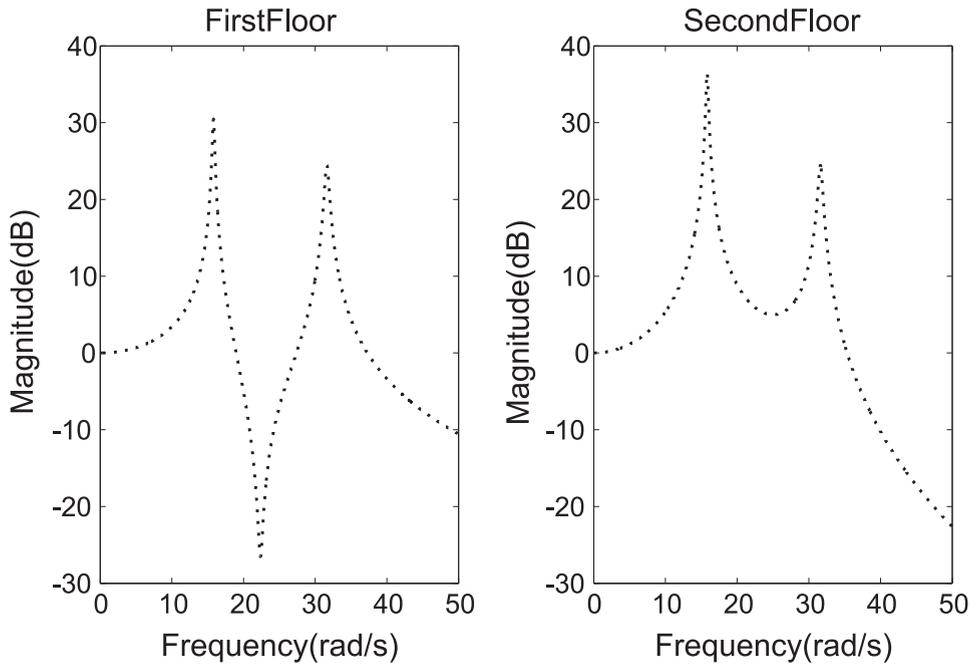

**Fig. 4.** FRF for the $N=2$ floors case without any TMD installed.

The CRO-SL with the previously defined parameters has been applied to solve two different application examples, consisting of designing and locating $M=N$ TMDs for a $N=2$ and $N=4$ storey building. The TMDs can be placed on any floor to damp any vibration mode. The FRF between the acceleration of each floor ($y_j$) and the acceleration of the ground ($a_g$) will be used to show the performance of the optimal design. The parameters for the $N=2$ floor building are the following: i) $k_1 = 1000$ N/m and $k_2 = 500$ N/m ii) $m_1 = 2$ kg and $m_2 = 1$ kg and iii) $\xi_s = 0.01$. With these parameters, the natural frequencies and damping are $\omega_1 = 15.811$ rad/s, $\omega_2 = 31.623$ rad/s and $\xi_1 = \xi_2 = 0.010$. The FRF between the acceleration of each floor ($y_j$) and the acceleration of the ground ($a_g$) is shown in Fig. 4 (without any TMD). The constraints for **x** are $\omega_{t,j} \in [0, 50]$ rad/s, $\xi_{t,j} \in [0, 0.3]$, $m_{t,j} \in [0, 0.05]$ kg and $fb_j \in \{1, 2\}$.

On the other hand, the parameters for the $N=4$ floor building problem are the following: i) $k_1 = 2000$ N/m, $k_2 = 1500$ N/m, $k_3 = 1000$ N/m and $k_4 = 500$ N/m, ii) $m_1 = m_2 = m_3 = 2$ kg and $m_4 = 1$ kg and iii) $\xi_s = 0.01$. With these parameters, the natural frequencies and damping are $\omega_1 = 10.608$ rad/s, $\omega_2 = 24.380$ rad/s, $\omega_3 = 34.538$ rad/s, $\omega_4 = 48.479$ rad/s, $\xi_1 = 0.020$, $\xi_2 = 0.011$ and $\xi_3 = \xi_4 = 0.010$. The FRF between the acceleration of each floor ($y_j$) and the acceleration of the ground ($a_g$) is shown in Fig. 5 (without any TMD). In this case, the restrictions for **x** are $\omega_{t,j} \in [0, 50]$ rad/s, $\xi_{t,j} \in [0, 0.3]$, $m_{t,j} \in [0, 0.05]$ kg and $fb_j \in \{1, 2, 3, 4\}$.

### 4.1. Results

Table 2 shows the results obtained by the proposed CRO-SL, compared to different alternative algorithms. Specifically, all the algorithms that form the substrate layers in the CRO-SL approach has been tried on their own, with the same number of function evaluations, in order to show how the competitive co-evolution process promoted by the CRO-SL is positive to obtain better solutions for the TMD design and location problem. In Table 2 it can be seen how the CRO-SL obtains the best performance, both in the $N=2$ and $N=4$ cases. In the $N=2$ case, the differences among different methods are small, with the HS as the second best approach, and the 2Px and MPx crossover quite close behind. In this case, it seems that the DE and GM exploration patterns works worse than the other search procedures considered. In the case of $N=4$, the differences are larger. The proposed CRO-SL approach obtains the best result, and in this case DE operator also obtains a very good solution, close to the best obtained by the CROSL. The Mpx is the third best approach in this instance, whereas the Gaussian mutation and HS operators seem to work worse in this harder problem.

The best result obtained by the CRO-SL in the case $N=2$ is the following:

$$\boldsymbol{\Omega}_t = [22.6586, 14.9481] \text{ rad/s}, \quad \boldsymbol{\Xi}_t = [0.2939, 0.1149], \quad \mathbf{M}_t = [0.0473, 0.0500] \text{ kg}, \quad \mathbf{FB} = [2, 2]. \tag{12}$$

Regarding this best solution (see Fig. 6), note the following: i) the two TMDs are located in the second floor, ii) the first

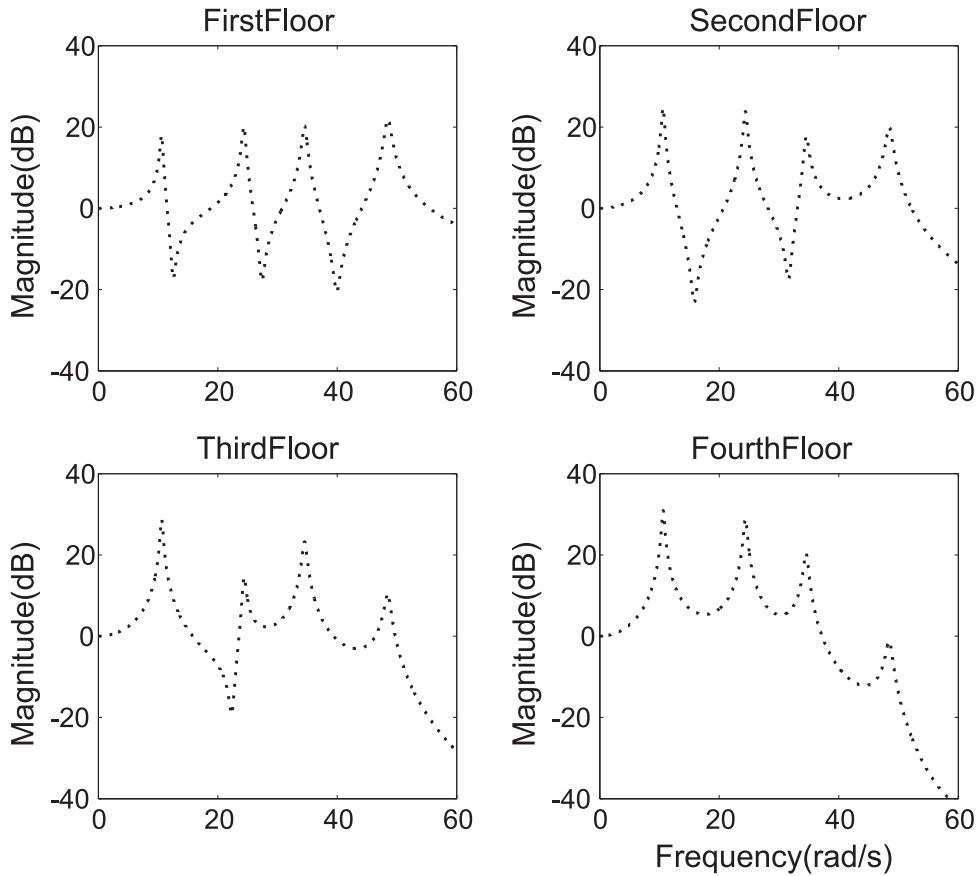

**Fig. 5.** FRF for the $N=4$ floors case without any TMD installed.

**Table 2**
Comparison of the results obtained in the two case-studies taken into account ($N=2$ and $N=4$) with different algorithms, in terms of the fitness function considered (Eq. (10)).

|  | 2 floors | | 4 floors | |
| --- | --- | --- | --- | --- |
|  | Min | Mean | Min | Mean |
| CRO-SL | 8.4348 | 8.5773 | 7.7746 | 7.8747 |
| HS | 8.4728 | 8.5786 | 8.848 | 9.4393 |
| DE | 9.5405 | 10.1129 | 7.8831 | 7.9833 |
| 2Px | 8.5306 | 8.627 | 8.9341 | 8.9897 |
| GM | 8.9914 | 9.162 | 10.3464 | 11.3341 |
| MPx | 8.7337 | 8.797 | 8.4458 | 8.9154 |

TMD is syntonized to the first vibration mode and iii) the natural frequency of the second TMD is between the first and second vibration mode of the structure. It should be remarked that the maximum of the FRFs without TMDs is located in the second floor-first vibration mode (36.5 dB). The maximum of the FRFs with TMDs is located in the second vibration mode, where the amplitudes are 18.4 and 18.6 dB for the first and second floor, respectively. Therefore, the second TMD is syntonized to damp both vibration modes and to level the amplitude of the second vibration mode at both floors. Fig. 7 (a) shows the evolution of the best solution found by the CRO-LS algorithm for the $N=2$ floors case. Note how the convergence of the proposed approach to the optimal solution is fast, in around 200 iterations. Note also that the CRO-SL is a co-evolution algorithm which evolves different exploration patterns in the substrate layers. The question is how to evaluate what is the substrate layer that contributes the most to the search in this problem of TMDs location and design. To clarify this point, Fig. 7 (b) shows the ratio of times that every substrate generates the best larva per generation in the CRO-SL approach ($N=2$ floors case). It indicates that the MPx crossover and the 2-points crossover are the two exploration operators that contribute the most to the CRO-SL search. The HS substrate seems to contribute to the CRO-SL search as well, and the rest of operators (substrates) contribute very little, and only in the earliest stages of the algorithm. This behaviour is consistent to the performance of the different algorithms run on their own, as shown in Table 2, where it was shown that the DE and GM searchers do not work well in this problem.

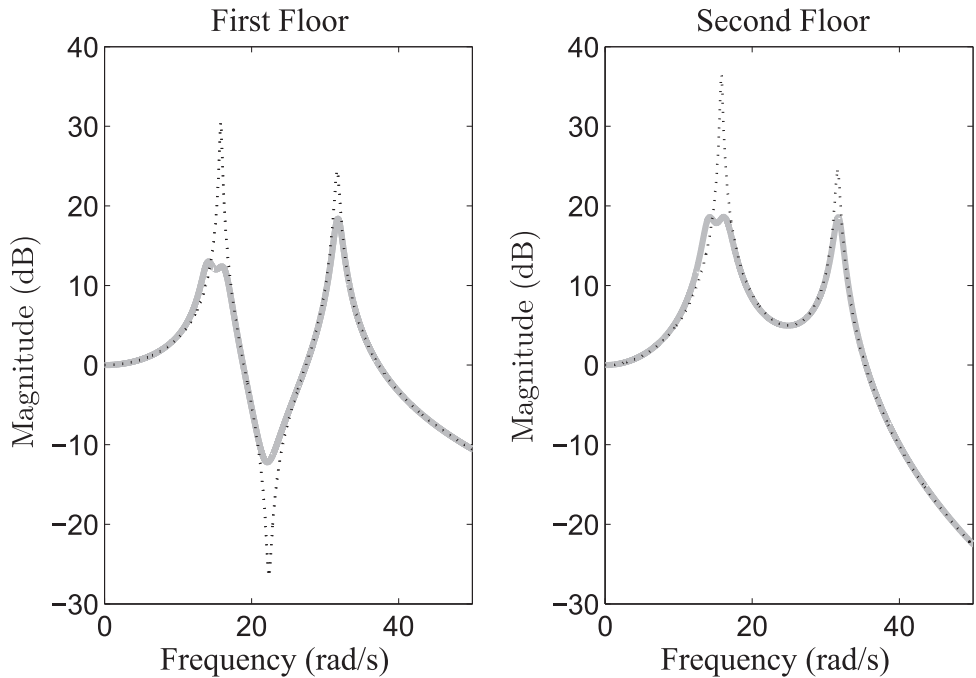

**Fig. 6.** FRF for the $N=2$ case, with the optimal position and design for the TMDs using the CRO-SL algorithm. ((⋯)-black) without TMDs and ((—)-gray) with TMDs.

The example of TMD design and location with $N=4$ floors is more complex as optimization algorithm, since the search space is much larger than in the $N=2$ problem. In order to better motivate this application example, two different evaluations are carried out: first, a free-location of the 4 TMDs and their parameters, for $N=4$. Second, in order to compare this solution, we consider the case of an intuitive solution in which the four TMDs are located in the top floor, and only the rest of their parameters are sought with the proposed CRO-SL. The best solutions obtained by the CRO-SL in these cases are the following:

$$\boldsymbol{\Omega}_t = [9.8264, 10.5978, 21.3608, 31.8252] \text{ rad/s}, \quad \boldsymbol{\Xi}_t = [0.0985, 0.1070, 0.2398, 0.3000],$$
$$\mathbf{M}_t = [0.0500, 0.0500, 0.0500, 0.0500] \text{ kg}, \quad \mathbf{FB} = [4, 4, 4, 1]. \tag{13}$$

in the first case (free TMD locations), and

$$\boldsymbol{\Omega}_t = [9.8887, 10.3113, 27.8916, 46.8533] \text{ rad/s}, \quad \boldsymbol{\Xi}_t = [0.3000, 0.1107, 0.1731, 0.0055],$$
$$\mathbf{M}_t = [0.0500, 0.0500, 0.0500, 0.0500] \text{ kg}, \quad \mathbf{FB} = [4, 4, 4, 4], \tag{14}$$

for the second evaluation problem in which TMD locations are pre-set.

The results of the four story building model ($N=4$) with the TMD optimal configuration of Eqs. (13) and (14) are shown in Fig. 8. The maximum of the FRFs is located in the fourth floor-first vibration mode without any TMD (30.9 dB). The maximum of values of the FRFs, when the four TMDs are located as Eq. (13), are in the third floor-third vibration mode, fourth floor-third vibration mode and fourth floor-fourth vibration mode (approximately equal to 17.8 dB). The maximum values of the FRFs, when the four TMDs are located as Eq. (14), are in fourth floor-first vibration mode, fourth floor-third vibration, third floor-third vibration mode and first floor-fourth vibration mode (approximately equal to 20.2 dB).

The following conclusions can be deduced from the comparison between both optimal solutions: i) both of them place two TMDs on fourth floor and their natural frequencies are close to the first vibration mode, ii) although both solutions place another TMD on fourth floor to reduce the second vibration mode, the natural frequency of the TMD corresponding to the first case (free TMD locations) is between first and vibration mode, which implies a better reduction in the first vibration mode with a less damping ratio, and iii) although it might seem unlikely, the last TMD is placed on first floor when there is freedom to locate TMDs, which improves the damping performance in the third and fourth vibration comparing with the design of all TMDs on the fourth floor. Therefore, although both evaluation problems show that the amplitude of the FRFs is reduced in all the vibration modes and in all the floors, the first case (free TMD locations in the CRO-SL) produces better results. Note that this illustrates an example where the optimum location is not the most obvious solution.

Fig. 9 (a) shows the evolution of the best solution found by the CRO-LS algorithm in this problem of TMD design and location for the case $N=4$ floors. Regarding the ratio of times that every substrate generates the best larva per generation in the CRO-SL approach, Fig. 9 (b) shows it for this problem. It indicates that the MPx crossover and the 2-points crossover are again the two exploration operators that contribute the most to the CRO-SL search, but in this problem the DE substrate seems to contribute more than in the previous case, whereas the HS and Gaussian substrates barely contribute to obtain the best solutions in each iteration of the CRO-SL. This also coincides to the results reported in Table 2.

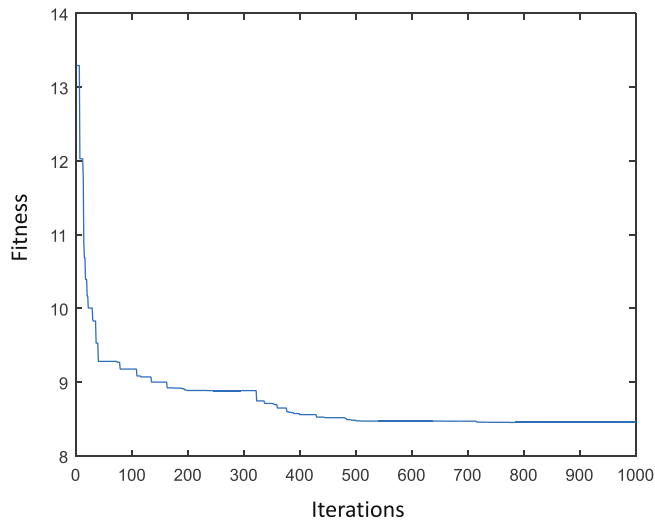

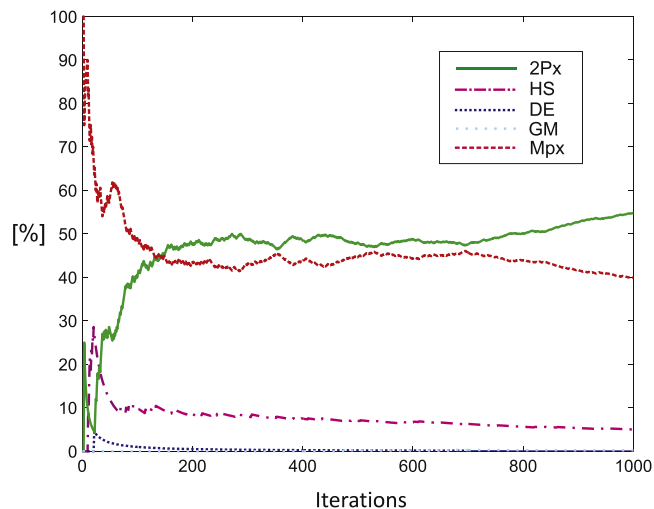

**Fig. 7.** Evolution of the best solution within the CRO-SL and ratio of times that each substrates produces the best larva in each iteration of the algorithm, for the $N=2$ floors TMDs location problem; (a) Best evolution; (b) Competitive ratio of the best solution found in each generation.

The results reported in this paper show that the CRO-SL is able to obtain excellent results for problems of TMD tuning and location, improving other meta-heuristics algorithms in this hard optimization problem in structures engineering.

## 5. Experimental implementation

The experimental set-up consists of a $N=2$ storey building. The parameters for the $N=2$ floor building are obtained through experimental identification of the scale model shown in Fig. 10 (a), instaled on a sliding table. For the experimental identification methodology, three accelerometers (MMF-KS76C, with 100 mV/g of sensitivity) attached to the ground, first floor and second floor, are connected to SIRIUS-HD-16xSTGS datalogger. The response under a soft impact applied to the sliding table is postprocessed using the Modal Testing tool available in DeweSoft X -DSA SP5. Sampling rate was set to 1000 Hz and the FRFs in Fig. 10 (c)-gray are obtained after averaging 5 impacts.

Values for the stiffness, mass and damping are $k_1 = 1111.8$ N/m, $k_2 = 389.1$ N/m, $m_1 = 2.14$ kg, $m_2 = 1.88$ kg and $\xi_s = 0.006$.

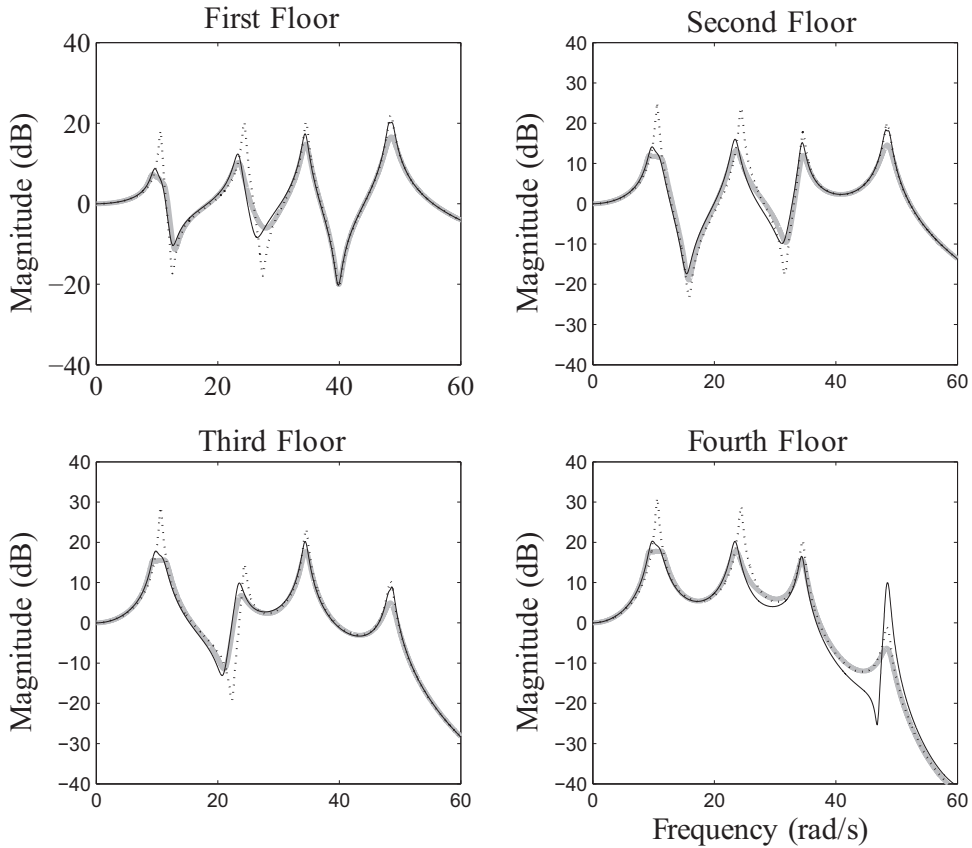

**Fig. 8.** FRF for the $N=4$ case, with the optimal position and design for the TMD using the CRO-SL algorithm. ((···)-black) without TMDs, ((−)-gray) with TMDs located at **FB**=[4,4,4,1] and ((−)-black) with TMDs located at **FB**=[4,4,4,4].

With these parameters, the natural frequencies and damping are $\omega_1 = 11.842$ rad/s, $\omega_2 = 27.733$ rad/s and $\xi_1 = \xi_2 = 0.006$. Fig. 10 (c) shows a very good agreement between computational and experimental FRFs, revealing that the hypothesis of proportional damping matrix can work well in practical engineering application.

The laboratory TMDs comprize (see Fig. 10 (b))) a cantilever leaf spring with adjustable length mounted in the frame. At the end of the cantilever the magnet is in close proximity to an aluminum plate, also supported in the frame. The effective moving mass depends not only on the masses (and magnet) placed at the end of the cantilever but also on the length of the leaf, that has been adjusted to obtain the tuning frequency of the TMD. The damping coefficient is adjusted moving closer or away the aluminum plate with regards to the magnet. This set-up, whith 3 mounting frames, is ready to undergo any of the logical solution of the problem, which are both TMD in the upper floor or each one on a floor.

In order to better motivate this application example (like $N=4$ example), two different evaluations are carried out: first, a free-location of the two TMDs and their parameters, for $N=2$. Second, in order to compare this solution, we consider the case of an intuitive solution in which the two TMDs are located in the top floor, and only the rest of their parameters are sought with the proposed CRO-SL. The best solutions obtained by the CRO-SL in these cases are the following:

$$\Omega_t = [23.3822, 11.3105] \text{ rad/s}, \quad \Xi_t = [0.2000, 0.1344], \quad \mathbf{M}_t = [0.100, 0.100] \text{ kg}, \quad \mathbf{FB} = [1, 2], \quad g(\mathbf{x}) = 7.5033, \quad (15)$$

in the first case (free TMD locations), and

$$\Omega_t = [11.3408, 26.6638] \text{ rad/s}, \quad \Xi_t = [0.1852, 0.0460], \quad \mathbf{M}_t = [0.0100, 0.0100] \text{ kg}, \quad \mathbf{FB} = [2, 2], \quad g(\mathbf{x}) = 9.8443, \quad (16)$$

for the second evaluation problem in which TMD locations are pre-set.

The results of the experimental set-up with the TMD optimal configuration of Eqs. (15) and (16) are shown in Fig. 11. Note that the intuitive solution in which the two TMDs are located in the top floor is worst than the one in which a first TMD placed on the second floor and tuned to the first mode and the second one is placed on the first floor. It is also noteworthy to realize that the second TMD is not really tuned to the second mode but placed in a frequency between the one of the first and the second mode. Note the good agreement, regardless of the difficulty in adjusting the values of the moving mass and the damping coefficient obtained by the optimization algorithm (i.e., the experimental values of the functions, $g(\mathbf{x})$, are approximately the same).

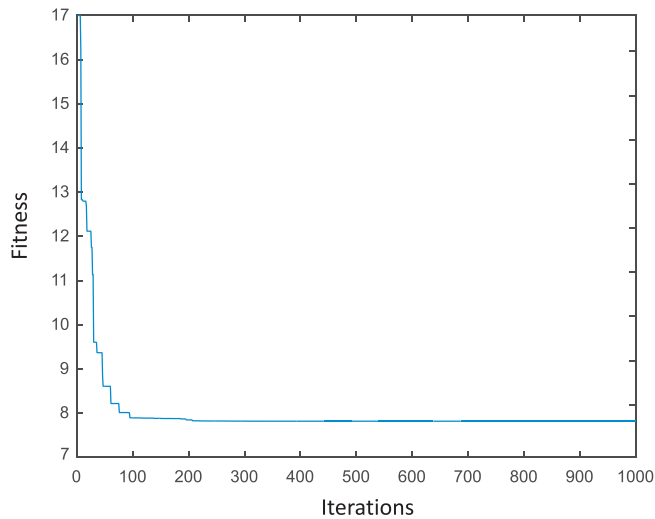

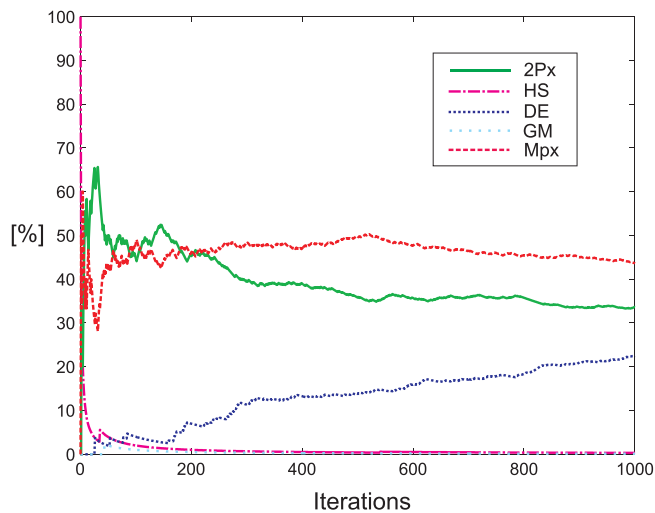

**Fig. 9.** Evolution of the best solution within the CRO-SL and ratio of times that each substrates produces the best larva in each iteration of the algorithm, for the $N=4$ floor TMD location problem; (a) Best evolution; (b) Competitive ratio of the best solution found in each generation.

## 6. Conclusions

In this paper, a novel co-evolution meta-heuristic to solve the design and location of Tuned Mass Dampers (TMDs) for structures subjected to earthquake ground motions has been proposed. Specifically, the Coral Reefs Optimization algorithm with Substrate Layer (CRO-SL) has been introduced for this problem. The CRO-SL algorithm collects together in a single method the exploration capabilities of different meta-heuristics search, and make them compete, by including them as different *substrates* in the algorithm. In order to apply the CRO-SL in a problem of optimal TMD tuning and location, a modification of the generalized framework presented in [28] has been used to formulate a $N$ floor building where $M$ TMDs must be installed. The proposed modification allows the optimization algorithm deciding the position of each TMD, in such a way that a TMD can be placed at any floor to damp any vibration mode. We have shown that the CRO-SL is an excellent approach to solve optimization problems related to the design and optimal location of TMDs in structures, by solving two case studies of TMD location in two building models with two and four floors, respectively, within a low computation time. Finally, an experimental set-up has been made to test the algorithm by using a model of the two floors building. These experimental results show that: i) the generalized framework proposed herein can work well in practical engineering applications, ii) the mass, stiffness and damping values of each TMDs can be accurate tuned with the laboratory equipment used.

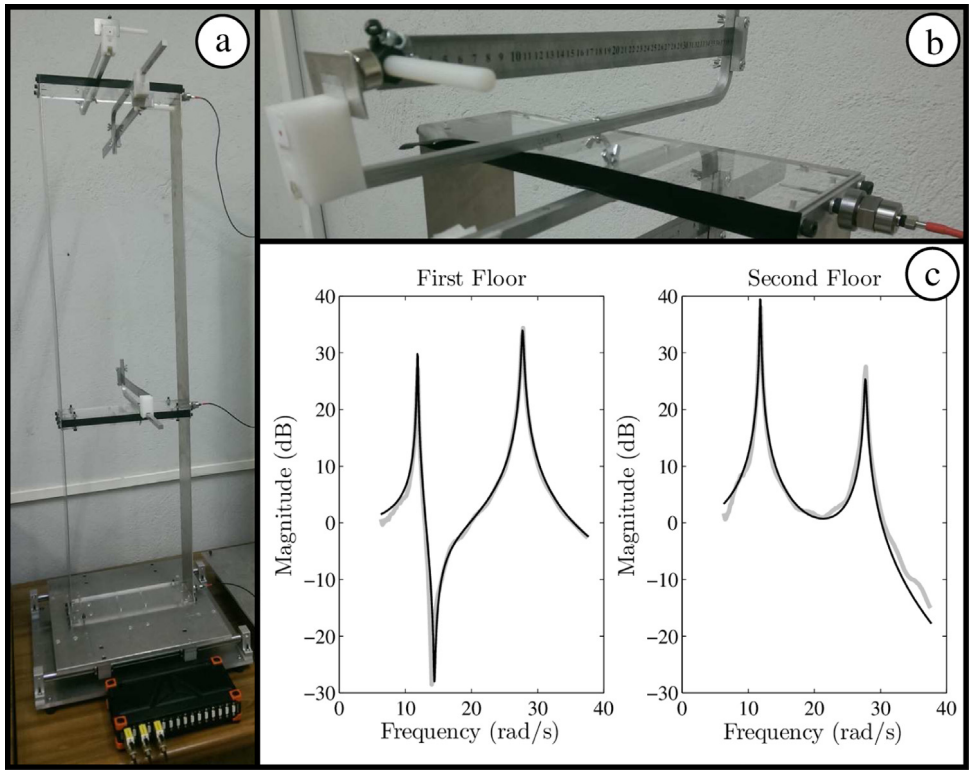

**Fig. 10.** Experimental set-up; a) $N=2$ floor building with TMD frames; b) Detail of the experimental TMD; c) Structure Identification, where ((−)-black) shows the FRF of the identified model, ((−)-gray) shows the experimental FRF.

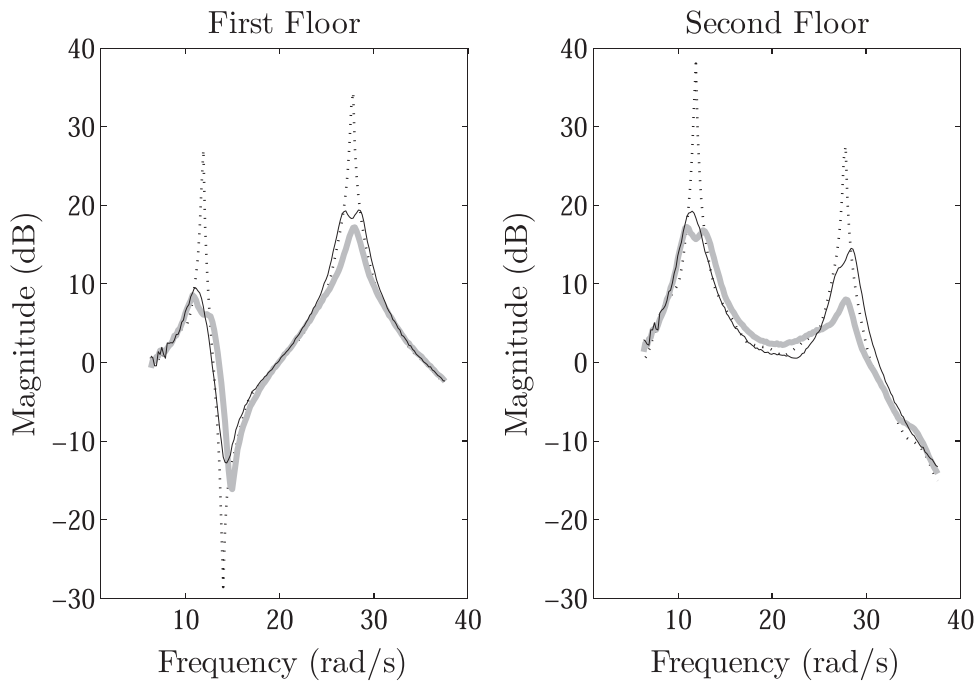

**Fig. 11.** Experimental FRF for the $N=2$ case, with the optimal position and design for the TMDs using the CRO-SL algorithm. ((⋯)-black) without TMDs, ((−)-gray) with TMDs located at **FB**=[2,1] and (((−)-black) with TMDs located at **FB**=[2,2].

**Acknowledgements**

This work has been partially supported by the Spanish Government under the projects numbers TIN2014-54583-C2-2-R, BIA2011-28493, DPI2013-47441 and BIA2014-59321, and by Comunidad Autónoma de Madrid, under project number S2013ICE-2933_02.

**References**

[1] M.P. Saka, O. Hasancebi, Z.W. Geem, Metaheuristics in structural optimization and discussions on harmony search algorithm, *Swarm Evolut. Comput.* 28 (2016) 88–97.
[2] F. Glover, G.A. Kochenberg (Eds.), Handbook of Metaheuristics, Kluwer Academic Publisher, New York, USA, 2003.
[3] A.E. Eiben, J.E. Smith, *Introduction to Evolutionary Computing*, Springer-Verlag, Berlin, 2003.
[4] S. Rajeev, C.S. Krishnamoorthy, Discrete optimization of structures using genetic algorithms, *ASCE J. Struct. Eng.* 118 (1992) 1233–1250.
[5] C.K. Soh, J. Yang, Fuzzy controlled genetic algorithm search for shape optimization, *ASCE J. Comput. Civil. Eng.* 10 (1996) 143–150.
[6] V. Togan, A.T. Daloglu, Optimization of 3D trusses with adaptive approach in genetic algorithms, *Eng. Struct.* 28 (7) (2006) 1019–1027.
[7] G. Yan, L. Zhou, Impact load identification of composite structure using genetic algorithms, *J. Sound Vib.* 319 (3-5) (2009) 869–884.
[8] H.L. Simonetti, V.S. Almeida, L. de Oliveira Neto, A smooth evolutionary structural optimization procedure applied to plane stress problem, *Eng. Struct.* 75 (2014) 248–258.
[9] K. Deb, Optimal design of a welded beam via genetic algorithms, *AIAA J.* 29 (11) (1991) 2013–2015.
[10] J. Kennedy, R.C. Eberhart, Particle swarm optimization, in: Proceedings of IEEE International Conference on Neural Networks, vol. IV, 1995, 1942–1948.
[11] A. Kaveh, A. Zolghadr, Democratic PSO for truss layout and size optimization with frequency constraints, *Comput. Struct.* 130 (2014) 10–21.
[12] J.J. Schutte, A.A. Groenwold, Sizing design of truss structures using particle swarms, *Struct. Multidiscip. Optim.* 23 (4) (2003) 261–269.
[13] Z.W. Geem, J.H. Kim, G.V. Loganathan, A new heuristic optimization algorithm: harmony search, *Simulation* 76 (2) (2001) 60–68.
[14] R.V. Rao, V.J. Savsani, D.P. Vakharia, Teaching-learning-based optimization: *a novel optimization method for continuous non-linear large scale problems*, *Inf. Sci.* 183 (1) (2011) 1–15.
[15] R.V. Rao, V.J. Savsani, D.P. Vakharia, Teaching-learning-based optimization: a novel method for constrained mechanical design optimization problem, *Comput.-Aided Des.* 43 (2011) 303–315.
[16] S.O. Degertekin, M.S. Hayalioglu, Sizing truss structures using teaching-learning based optimization, *Comput. Struct.* 119 (2013) 177–188.
[17] A. Kaveh, V.R. Mahdavi, Optimal design of structures with multiple natural frequency constraints using a hybridized BB-BC/Quasi-Newton algorithm, *Period. Polytech. - Civil. Eng.* 57 (2013) 1–12.
[18] A. Kaveh, V.R. Mahdavi, Colliding bodies optimization: a novel meta-heuristic method, *Comput. Struct.* 139 (2014) 18–27.
[19] A. Kaveh, M. Khayatazad, A new meta-heuristic method: ray optimization, *Comput. Struct.* 112 (2012) 283–294.
[20] A. Kaveh, S.A. Talatahari, Novel heuristic optimization method: charged system search, *Acta Mech.* 213 (2010) 267–289.
[21] S. Salcedo-Sanz, C. Camacho-Gómez, D. Molina, F. Herrera, A coral reefs optimization algorithm with substrate layers and local search for large scale global optimization, in: Proceedings of the IEEE Conference on Evolutionary Algorithms, Vancouver, Canada (2016), 1–8.
[22] M. Symans, M. Constantinou, Semi-active control systems for seismic protection of structures: a state-of-the-art review, *Eng. Struct.* 21 (6) (1999) 469–487.
[23] J.P. Den-Hartog, *Mechanical Vibrations*, McGraw-Hill, New York, 1956.
[24] M. Abe, T. Igusa, Tuned mass dampers for structures with closely spaced natural frequencies, *Earthq. Eng. Struct. Dyn.* 24 (1995) 247–261.
[25] R. Greco, A. Lucchini, G. Marano, Robust design of tuned mass dampers installed on multi-degreeof-freedom structures subjected to seismic action, *Eng. Optim.* 47 (8) (2015) 1009–1030.
[26] L. Fleck-Fade-Miguel, L. Fleck-Fadel-Miguel, R. Holdorf-Lopez, Robust design optimization of friction dampers for structural response control, *Struct. Control Health Monit.* 332 (9) (2014) 6044–6062.
[27] N. Debnath, S. Deb, A. Dutta, Frequency band-wise passive control of linear time invariant structural systems with h-infinity optimization, *J. Sound Vib.* 332 (23) (2013) 6044–6062.
[28] A. Mohtat, E. Dehghan-Niri, Generalized framework for robust design of tuned mass damper systems, *J. Sound Vib.* 330 (5) (2011) 902–922.
[29] J. Woodhouse, Linear damping models for structural vibration, *J. Sound Vib.* 215 (3) (1998) 547–569.
[30] W.K. Gawronski, Advanced Structural Dynamics and Active Control of Structures, Mechanical Engineering Series, Springer Link, 2004.
[31] S. Salcedo-Sanz, J. Del Ser, I. Landa-Torres, S. Gil-López, J.A. Portilla-Figueras, The coral reefs optimization algorithm: a novel metaheuristic for efficiently solving optimization problems, *Sci. World J.* 2014 (2014). (Article ID: 739768).
[32] S. Salcedo-Sanz, D. Gallo-Marazuela, A. Pastor-Sánchez, L. Carro-Calvo, J.A. Portilla-Figueras, L. Prieto, Offshore wind farm design with the coral reefs optimization algorithm, *Renew. Energy* 63 (2014) 109–115.
[33] S. Salcedo-Sanz, A. Pastor-Sánchez, J. Del Ser, L. Prieto, Z.W. Geem, A coral reefs optimization algorithm with harmony search operators for accurate wind speed prediction, *Renew. Energy* 75 (2015) 93–101.
[34] Z. Yang, T. Zhang, D. Zhang, A novel algorithm with differential evolution and coral reef optimization for extreme learning machine training, *Cogn. Neurodyn.* 10 (1) (2016) 73–83.
[35] S. Salcedo-Sanz, C. Casanova-Mateo, A. Pastor-Sánchez, M. Sánchez-Girón, Daily global solar radiation prediction based on a hybrid coral reefs optimization - extreme learning machine approach, *Sol. Energy* 105 (2014) 91–98.
[36] I.G. Medeiros, J.C. Xavier-Júnior, A.M. Canuto, Applying the coral reefs optimization algorithm to clustering problems, in: Proceedings of the International Joint Conference on Neural Networks (IJCNN), 2015, pp. 1–8.
[37] M. Li, C. Miao, C. Leung, A coral reef algorithm based on learning automata for the coverage control problem of heterogeneous directional sensor networks, *Sensors* 15 (2015) 30617–30635.
[38] M.J. Vermeij, Substrate composition and adult distribution determine recruitment patterns in a Caribbean brooding coral, *Mar. Ecol. Progress. Ser.* 295 (2005) 123–133.
[39] S. Salcedo-Sanz, J. Muñoz-Bulnes, M. Vermeij, New coral reefs-based approaches for the model type selection problem: a novel method to predict a nation's future energy demand, *Int. J. Bio-inspired Comput.* (2017). in press.
[40] S. Salcedo-Sanz, A review on the coral reefs optimization algorithm: new development lines and current applications, Progress. Artif. Intell. 2017, http://dx.doi.org/10.1007/s1374801601042, in press.
[41] D. Kirpatrick, C.D. Gerlatt, M.P. Vecchi, Optimization by simulated annealing, *Science* 220 (1983) 671–680.